\documentclass[10pt,twocolumn,aps,pra,showpacs,superscriptaddress,floatfix,nofootinbib]{revtex4-1}

\usepackage{graphicx}
\usepackage{amssymb}
\usepackage{amsmath}
\usepackage{color}
\usepackage{psfrag}
\usepackage{epsfig}
\usepackage{bbm}
\usepackage{bm}
\usepackage{hyperref}
\usepackage[normalem]{ulem}
\usepackage{amsthm}
\usepackage{epstopdf}

\newcommand{\ket}[1]{| #1 \rangle}
\newcommand{\bra}[1]{\langle #1 |}
\newcommand{\braket}[1]{\langle  #1 \rangle}

\newcommand{\beq}{\begin{eqnarray}}
\newcommand{\eeq}{\end{eqnarray}}

\newcommand{\mean}[1]{\langle #1 \rangle}
\newcommand{\id}{\openone}
\newcommand{\SCHSH}{\mathcal{S}_{\mbox{\tiny CHSH}}}
\newcommand{\A}{\mathcal{A}}
\newcommand{\B}{\mathcal{B}}
\newcommand{\M}{\mathcal{M}}
\newcommand{\GMA}{\{M_{a|x}\}_{x,a}}
\newcommand{\MAA}{\{A_{a|x}\}_{x,a}}
\newcommand{\MAB}{\{B_{b|y}\}_{y,b}}
\newcommand{\SAA}{\{\sigma_{b|y}\}_{b,y}}
\newcommand{\SAB}{\{\rho_{a|x}\}_{a,x}}
\newcommand{\sB}{{\mbox{\tiny B}}}
\newcommand{\rab}{{\rho_{\mbox{\tiny AB}}}}

\definecolor{nblue}{rgb}{0.2,0.2,0.7}
\definecolor{ngreen}{rgb}{0.2,0.6,0.2}
\definecolor{nred}{rgb}{0.8,0.2,0.2}
\definecolor{nblack}{rgb}{0,0,0}
\definecolor{hotmagenta}{rgb}{1.0, 0.11, 0.81}

\newcommand{\blu}{\color{nblue}}

\newcommand{\revision}[1]{{#1}}

\DeclareMathOperator{\tr}{tr}

\theoremstyle{definition}

\begin{document}
\title{Natural framework for device-independent quantification of quantum steerability, measurement incompatibility, and self-testing}

\author{Shin-Liang Chen}
\email{shin.liang.chen@phys.ncku.edu.tw}
\affiliation{Department of Physics, National Cheng Kung University, Tainan 701, Taiwan}
\author{Costantino Budroni}
\email{costantino.budroni@uni-siegen.de}
\affiliation{Naturwissenschaftlich-Technische Fakult\"at, Universit\"at
 Siegen, Walter-Flex-Str.\ 3, D-57068 Siegen, Germany}
\author{Yeong-Cherng Liang}
\email{ycliang@mail.ncku.edu.tw}
\affiliation{Department of Physics, National Cheng Kung University, Tainan 701, Taiwan}
\author{Yueh-Nan Chen}
\email{yuehnan@mail.ncku.edu.tw}
\affiliation{Department of Physics, National Cheng Kung University, Tainan 701, Taiwan}
\affiliation{Physics Division, National Center for Theoretical Sciences, Hsinchu 300, Taiwan}

\date{ \today}

\begin{abstract}
We introduce the concept of {\em assemblage moment matrices}, i.e., a collection of matrices of expectation values, each associated with a conditional quantum state obtained in a steering experiment. We demonstrate how it can be used for quantum \revision{states and measurements} characterization in a device-independent manner, i.e., without invoking any assumption about the measurement or the preparation device. Specifically, we show how the method can be used to lower bound the steerability of an underlying quantum state directly from the observed correlation between measurement outcomes. Combining such device-independent quantifications with earlier results established by Piani and Watrous [Phys. Rev. Lett. {\bf 114}, 060404 (2015)], our approach immediately provides a device-independent lower bound on the generalized robustness of entanglement, as well as the usefulness of the underlying quantum state for a type of subchannel discrimination problem.  In addition, by proving a quantitative relationship between steering robustness and the recently introduced incompatibility robustness, our approach also allows for a device-independent quantification of the incompatibility between various measurements performed in a Bell-type experiment. Explicit examples where such bounds provide a kind of self-testing of the performed measurements are provided.
\end{abstract}

\maketitle

A central feature of quantum theory is that certain properties of physical systems are complementary, and thus cannot be simultaneously determined by the act of measurements. A well-known example of this is a system's position and momentum, which led \revision{\em Einstein-Podolksy-Rosen} (EPR)~\cite{EPR35} to question the completeness of quantum theory. In modern terminologies, measurements associated with such properties are dubbed incompatible~\cite{Wolf09}, or {\em nonjointly-measurable}~\cite{BuschBook}, and their existence 
forbids a classical interpretation of quantum probabilities, thus giving rise to many celebrated quantum phenomena: from uncertainty relations \cite{BLW14,UPrev,BLW13,BLWRev}, contextuality~\cite{KS67,Klyachko08,Cabello08,Liang:PRep}, to various forms of {\em quantum nonlocality}~\cite{Bell64,NLReview,wiseman2007}.

Among these, Bell nonlocality~\cite{Bell64,NLReview}---which reflects the failure of the intuitive notion of local causality~\cite{Bell04}---has attracted more and more attention from the physics community, thanks to the advent of quantum information~\cite{Nielsen2011} and also to the \revision{exquisite control that we now have} over various quantum systems (see, e.g.,~\cite{Rowe01,Shadbolt2012,Lanyon14,Poh15,Hensen15,Giustina15,Shalm15,Christensen15} and references therein). Likewise what has come to be known as EPR-steering~\cite{Schr35,wiseman2007,Cavalcanti09}---which concerns the peculiar feature of quantum theory that allows one to remotely {\em steer} the physical state of a distant party---has now found application both in entanglement witnessing~\cite{wiseman2007}, as well as in quantum key distribution~\cite{Branciard12,Kogias16,Xiang16}, when only a subset of the parties have trusted devices.

In the extreme scenario when {\em none} of the devices are to be trusted, analysis is carried out within the so-called {\em device-independent} paradigm~\cite{Scarani12,NLReview}. In recent years, measurement statistics that manifest Bell nonlocality have been recognized as an indispensable resource in device-independent quantum information processing tasks, such as quantum key distributions~\cite{Barrett05,Acin07,Reichardt13,Vazirani14,Miller:1402.0489}, randomness extraction~\cite{Chung14}, and randomness expansion~\cite{Colbeck09,Pironio10,Colbeck11,Miller:1402.0489}. These {\em nonlocal} correlations have also been applied in the robust characterization of quantum resources, e.g., in the self-testing~\cite{Mayers04,McKague12} of quantum apparatus, in dimension witnessing~\cite{Brunner08}, as well as in versatile entanglement witnessing~\cite{Bancal11,Brunner12} and entanglement quantification~\cite{Moroder13,Toth15,Liang15} of quantum states.

Recent progress~\cite{Yang13,Yang14,Wu14,Pal14,Wang2016} on self-testing, however, has focused predominantly on the characterization of an unknown quantum state, paying relatively little attention to the measurements performed (see, however,~\cite{Bancal15,Supic:1601.01552}). In fact, powerful tools~\cite{Yang14,Bancal15} have been developed for the former, but an analog for the self-testing of general measurements have so far remained elusive. Here, we address this problem by considering a particular kind of moment matrices associated with unnormalized quantum states, namely, matrices of expectation values associated with the conditional states obtained in EPR-steering-type experiments.

Indeed, moment matrices have found applications in scenarios involving various levels of trust on the measurement devices: from the scenario where {\em all} devices are fully characterized~\cite{Shchukin05,Rigas06,Haseler08,Miranowicz09},  to the device-independent scenario~\cite{NPA,NPA2008,Moroder13} when {\em none} are trusted, as well as the intermediate steering scenario~\cite{pusey2014,Law2014,Kogias15,Sainz15} when only a {\em subset} of the parties have trusted devices. In particular, the moment matrices proposed by Navascu\'es-Pironio-Ac\'in (NPA)~\cite{NPA,NPA2008} and a variant proposed by Moroder {\em et al.}~\cite{Moroder13} are routinely used for various analyses in the device-independent setting---from the characterization of the set of quantum correlations (see also~\cite{Doherty08}) to the quantification~\cite{Moroder13,Toth15} of bipartite (multipartite) entanglement by (genuine) negativity~\cite{Vidal02}, and the computation of $k$-producible bounds~\cite{Curchod15} for the generation of device-independent witnesses for \revision{genuine multipartite entanglement~\cite{Bancal11,Pal11,Bancal12} (or more generally entanglement depth)~\cite{Liang15} etc.}

\begin{figure}[h!]
\emph{\includegraphics[width=7.5cm]{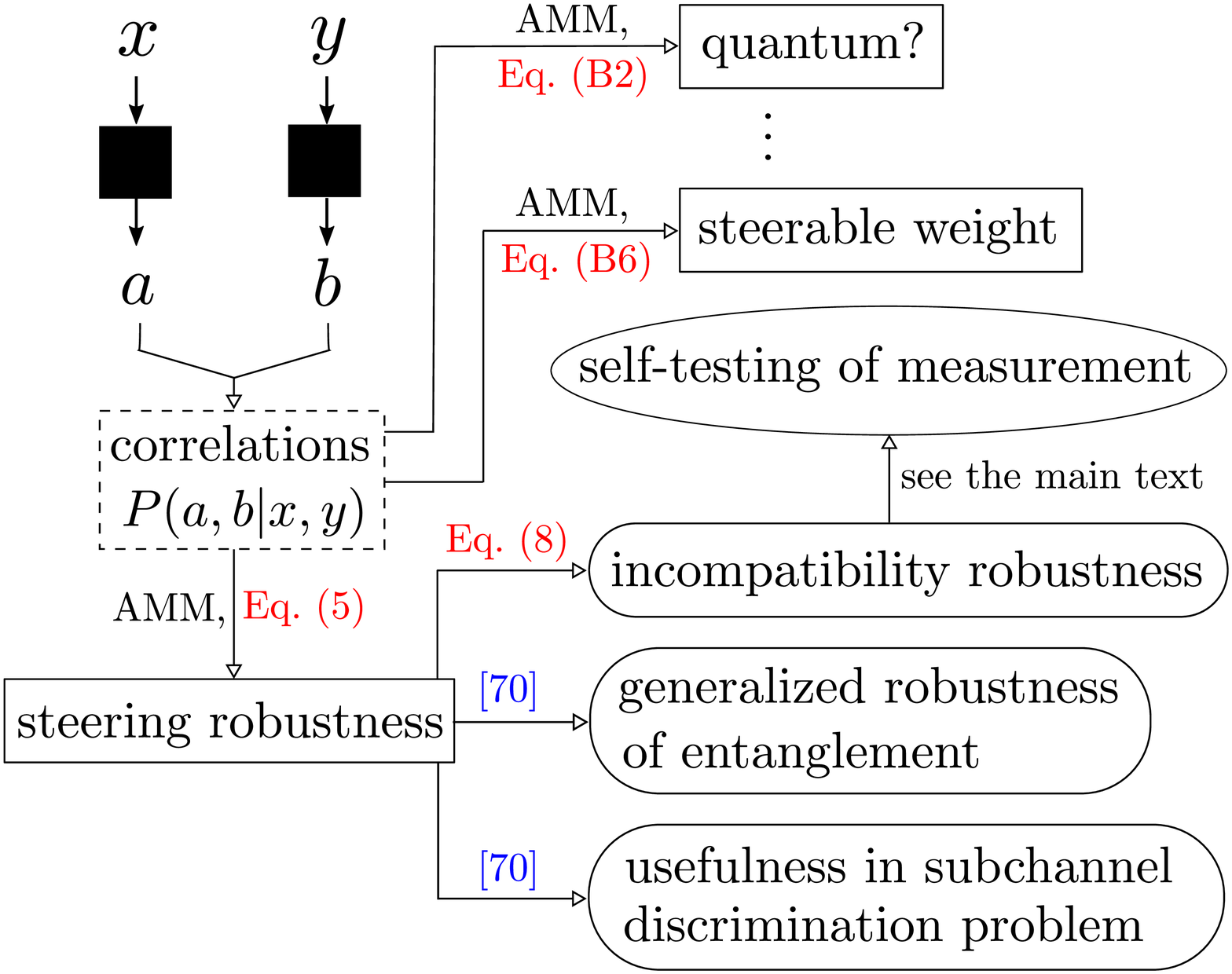} }
\caption{Schematic diagram showing how \revision{our {\em assemblage moment matrices} (AMM)} approach allows for a black-box estimation [i.e., using only $P(a,b|x,y)$] of various quantities of interest, such as steering robustness via Eq.~\eqref{relax_SR}, steerable weight via Eq.~(\ref{SDP:SW}), and incompatibility robustness of measurements via Eq.~\eqref{Eq:IR-SR}.}
\label{Fig1}
\end{figure}

As we shall demonstrate below, our moment matrices provide a natural framework for the device-independent quantification of steerability --- the extent to which an unknown quantum state can exhibit EPR steering. This, in turn, allows us to lower-bound other quantities of interest (see Fig.~\ref{Fig1}). For instance, by strengthening the recently established connection between steerability---quantified via steering robustness \cite{Piani15}---and incompatible measurements---quantified via incompatibility robustness \cite{Uola15}---our device-independent quantification of steerability immediately gives rise to a device-independent quantification of measurement incompatibility. The latter estimate, in turn, provides a figure of merit suited for the self-testing of measurements.

{\em Preliminary notions.---}
The generic setup that we shall consider is one that involves two spatially separated experimenters, called Alice and Bob, who, respectively, perform measurements labeled by $x\in\mathcal{X}$ and $y\in\mathcal{Y}$ on some shared quantum state $\rab$, each giving rise to measurements outcomes labeled by $a\in\mathcal{A}$ and $b\in\B$. (We denote throughout by $|\mathcal{S}|$ the cardinality of the set $\mathcal{S}$.) 

In quantum theory, a general measurement on a physical system is represented by a \revision{\em positive-operator-valued measure} (POVM) $\{M_{a|x}\}_{a}$, i.e., a collection of positive-semidefinite operators $M_{a|x}\succeq 0$ such that $\sum_a M_{a|x} = \openone$, with $\openone$ being the identity operator acting on the corresponding Hilbert space. Following~\cite{Piani15}, we refer to a collection of POVMs $\mathcal{M}=\{M_{a|x}\}_{x,a}$ as a {\it measurement assemblage}. Such a collection is said to be  {\it jointly measurable} \cite{BuschBook} 
if there exists an $|\A|^{|\mathcal{X}|}$ outcome POVM  $\{G_\lambda\}_\lambda$ (i.e., $G_\lambda \succeq 0$ and $\sum_\lambda G_\lambda = \openone$) such that each $M_{a|x}\in\GMA$ can be inferred from $G_\lambda$ via coarse graining, namely,
\begin{equation}\label{eq:JM}
	M_{a|x} = \sum_\lambda D(a|x,\lambda)\ G_\lambda\quad\forall\quad x,a,
\end{equation}
where $D(a|x,\lambda)\geq 0$ and $\sum_a D(a|x,\lambda) =1$.

Denoting Alice's and Bob's measurement assemblage, respectively, by $\MAA$ and $\MAB$, Born's rule dictates that the measurement outcomes appear according to the conditional probability distributions $P(a,b|x,y)=\tr[(A_{a|x}\otimes B_{b|y}) \rab]$. Moreover, whenever Alice obtains the outcome $a$ for the measurement $x$, Bob obtains the state $\varrho(a|x)$ with probability  $P(a|x)=\tr[(A_{a|x}\otimes \openone) \rab]$. 
 
In the context of EPR steering, it turns out to be more convenient to work with the unnormalized quantum states  $\rho_{a|x}:=P(a|x) \varrho(a|x) =  \tr_\text{A}[(A_{a|x}\otimes \openone) \rab]$, where $\tr_\text{A}$ denotes the partial trace over Alice's Hilbert space. In particular, any measurement assemblage $\MAA$ gives rise to a collection of unnormalized conditional quantum states $\{\rho_{a|x}\}_{a,x}$, termed a {\it state assemblage}~\cite{pusey2014} (henceforth \revision{abbreviated as an} {\em assemblage} when there is no risk of confusion). An important feature of any assemblage is that 
Bob's reduced state can be recovered as $\rho_{\mbox{\tiny B}} = \sum_a \rho_{a|x}$ for all $x$.

A given assemblage $\{\rho_{a|x}\}_{a,x}$ is said to admit a \revision{\em local-hidden-state} (LHS) model~\cite{wiseman2007} if there exists a collection of unnormalized states  $\{ \sigma_\lambda\}_\lambda$ such that $\tr[\sum_\lambda \sigma_\lambda]=1$ and~\cite{pusey2014}
\begin{equation}\label{eq:LHS}
	\rho_{a|x} = \sum_\lambda D(a|x,\lambda)\ \sigma_\lambda\quad\forall\quad a,x,
\end{equation}
with $D(a|x,\lambda)\geq 0$ and $\sum_a D(a|x,\lambda) =1$. If a LHS model exists for a given assemblage, Bob can always interpret each $\rho_{a|x}$ as coming from some preexisting states $\sigma_\lambda$, where only the classical probabilities are updated due to the information obtained by Alice from her measurement. A {\em steerable} assemblage is one that cannot be explained by any LHS model. Likewise, a quantum state $\rab$ is  {\em steerable} if \revision{it} can give rise to a steerable assemblage; such a state is necessarily entangled, but the converse is not true~\cite{wiseman2007}.

{\em The assemblage moment matrices.---} To connect  steerability and measurement incompatibility in the device-independent setting (and hence to allow for the possibility to self-test measurements), let us now \revision{briefly recall from \cite{Moroder13} the local mapping introduced by Moroder {\em et al.},} which we shall apply instead to a (not necessarily normalized) single-partite state $\rho$. Let $\chi$  be the moment matrix induced by the measurement assemblage $\M=\{M_{a|x}\}_{x,a}$ on $\rho$, that is, $\chi[\rho] = \Lambda(\rho)=\sum_n K_n \rho K_n^\dagger$ where $\Lambda:\mathcal{H}_S\rightarrow \mathcal{H}_{\bar{S}}$ is a completely positive map  with Kraus operators given by $K_n=\sum_i \ket{i}_{\bar S \: S\!}\bra{n} O_i$, while $\{\ket{i}\}_i$, $\{\ket{n}\}_n$ are, respectively, orthonormal basis for the \revision{output space $\bar{S}$ \& input space $S$}. In the above definition of the Kraus operators, $\{O_i\}_i := \{\openone\}\bigcup\M \setminus\{M_{|\A| |x}\}_x$ is the union of $\openone$ and all but the last-outcome POVM elements. Explicitly, $\chi[\rho]$ is thus a matrix of expectation values $\sum_{ij} \ket{i}\bra{j} \tr[\rho\, O_j^\dagger O_i]$ containing \revision{\em only} second and lower order moments. More generally, by applying a completely positive map $\Lambda$ with Kraus operators~\cite{Moroder13} $K_n = \sum_{i_1,\ldots,i_\ell} \ket{i_1,\ldots,i_\ell}_{\bar S \: S\!}\bra{n} O_{i_1},\ldots,O_{i_\ell}$  on $\rho$, we obtain a $\chi[\rho]$ whose entries are the moments with order $2\ell$ or lower --- henceforth, we shall refer to this as the level-$\ell$ moment matrix. Evidently, since $\id\in\{O_i\}_i$, one of the entries of $\chi[\rho]$ is the trace of $\rho$ which we shall, following~\cite{Moroder13}, denote by $\chi [\rho]_{\tr}=\tr(\rho)$.

Consider now the application of the above mapping ---defined via Bob's  local measurements $\{ B_{b|y}\}_{y,b}$---on Bob's assemblage $\{ \rho_{a|x}\}_{a,x}$. \revision{A straightforward} calculation shows that for each  $(a,x)$, we obtain a moment matrix containing entries $P(a,b|x,y)$ that are directly accessible from experimental observation, and open entries (unknowns) that are accessible {\em only if} we are given complete specification of the assemblage $\SAB$ as well as the measurement operators $\{B_{b|y}\}_{y,b}$ (see Appendix~\ref{App:AMM:Level1}). Following~\cite{Moroder13}, we decompose the moment matrix as 
\begin{equation}
\begin{split}
\chi[\rho_{a|x}] &= \chi[P,u]=\chi^{\rm fixed} (P) + \chi^{\rm open} (u)\\
&=\sum_{b,y} P(a,b|x,y) F_{abxy} + \sum_v u_v F_v,
\end{split}
\end{equation}
where each $F_{abxy}$ and each $F_u$ are boolean (Hermitian) matrices, and all \revision{(higher order)} moments that are not accessible  experimentally are collected as $\{u_v\}_v$.  Note that the entry $\chi [\rho_{a|x}]_{\tr}$ now gives the marginal distribution $P(a|x)$. The set $\{\chi[\rho_{a|x}]\}_{a,x}$ constitutes what we shall refer to as the {\em assemblage moment matrices} (AMM). One can see them as the result of first applying the local mapping of~\cite{Moroder13} at the first level on Alice's Hilbert space, but local mapping at the $\ell$ level on Bob's Hilbert space, and \revision{taking $|\A||\mathcal{X}|$ smaller submatrices with entries corresponding to only first order moments of Alice} (see Appendix~\ref{App:AMM} for further details about AMM). As with the moment matrices introduced in~\cite{Moroder13}, or in the NPA moment matrices~\cite{NPA}, since the AMM contain all the accessible moments $P(a,b|x,y)$, they can be used to characterize the set of quantum correlations, and to compute upper bounds on so-called Tsirelson bounds~\cite{Tsirelson1980} \revision{by solving a {\em semidefinite program}} (SDP)~\cite{BoydBook} (see Appendix~\ref{App:SDP}). In particular, as we increase the level $\ell$ of the AMM considered, we obtain a better characterization of the set of quantum correlations (and hence a tighter upper bound on the Tsirelson bound). For explicit examples of such computational results, see Appendix~\ref{App:TsirelsonBounds}.

{\em Device-independent quantification of steerability.---}The steerability of an assemblage $\{\rho_{a|x}\}_{a,x}$, and hence of a quantum state $\rab$ can be quantified, for instance, by the steerable weight~\cite{SNC14} or the \revision{\em steering robustness} (SR)~\cite{Piani15}, both efficiently computable---given the assemblage---through \revision{a SDP}. In particular, SR is quantitatively tied to the probability of success in the problem of subchannel discrimination when one is restricted to local measurements aided by \revision{a one-way} communication~\cite{Piani15}.  Explicitly, for any given assemblage $\{\rho_{a|x}\}_{a,x}$, its SR, which we denote by $\text{SR}(\{\rho_{a|x}\}_{a,x})$ is defined~\cite{Piani15} as the minimum $t$ such that $\{ (\rho_{a|x} + t \tau_{a|x})/(1+t) \}_{a,x}$ is \revision{unsteerable} for some state assemblage $\{ \tau_{a|x} \}_{a,x}$; it can be computed as:
\begin{subequations}\label{SR}
\begin{align} 
\min_{\{\sigma_\lambda\}} ~~&\left(\sum_{\lambda}\text{tr}\,\sigma_{\lambda}\right)-1 \label{Eq:SRObj}\\
\text{subject to }~~ &\sum_{\lambda}D(a|x,\lambda)\sigma_{\lambda}\succeq\rho_{a|x}\quad\forall~a,x\label{Eq:SRConstr1}\\
&\qquad\sigma_{\lambda}\succeq 0\qquad \forall ~\lambda\label{Eq:SRConstr2}
\end{align} 
\end{subequations}
where the sum above can be taken over all deterministic strategies  $D(a|x,\lambda)=\delta_{a,{\lambda_x}}$, with $\lambda=(\lambda_1,\ldots,\lambda_{|\mathcal{X}|})$, and $A \succeq B$ means that $A-B$ is positive semidefinite. 

The ability of Alice to steer Bob's assemblage when they share a quantum state $\rab$ can then be quantified by maximizing $\text{SR}(\{\rho_{a|x}\}_{a,x})$ over all possible assemblages   $\{\rho_{a|x}\}_{a,x}$ derivable from $\rab$ via Alice's measurements $\{A_{a|x}\}_{x,a}$, i.e., $\text{SR}_{\rm A\to B}(\rab)=\max_{\{\rho_{a|x}\}_{a|x}} \text{SR}(\{\rho_{a|x}\}_{a,x})$ such that $\rho_{a|x} =  \tr_\text{A}[(A_{a|x}\otimes \openone) \rab]$. The ability for Bob to steer Alice's assemblage can be analogously defined as $\text{SR}_{\rm B\to A}(\rab)=\max_{\{\sigma_{b|y}\}_{b|y}} \text{SR}(\{\sigma_{b|y}\}_{b,y})$ such that $\sigma_{b|y} =  \tr_\text{B}[(\openone\otimes B_{b|y}) \rab]$. Together, this allows one to define the steerability of a quantum state $\rab$ as SR$(\rab)=\max \{\text{SR}_{\rm A\to B}(\rab),\text{SR}_{\rm B\to A}(\rab)\}$.

In the device-independent setting, one {\em does not} have access to the assemblage $\{\rho_{a|x}\}_{a,x}$. Nonetheless, an estimate of  $\text{SR}(\{\rho_{a|x}\}_{a,x})$ and hence that of $\text{SR}(\rab)$ can still be achieved from the observed correlations $\vec{P}_{\mbox{\tiny obs}}=\{P_{\mbox{\tiny obs}}(a,b|x,y)\}$ by considering a relaxation of Eq.~\eqref{SR} using AMM. In fact, since ${\rm SR}(\SAB)$ does not increase under local channels \cite{Gallego15}, for any given level of the AMM, solving the SDP
\begin{subequations}\label{relax_SR}
\begin{align}
\min_{\{u_v\}} ~~&\left(\sum_{\lambda}\chi[\sigma_{\lambda}]_\text{tr}\right)-1 \label{Eq:SRObj-Relaxed}\\
\text{subject to }~~ &\sum_{\lambda}D(a|x,{\lambda})\chi[\sigma_{\lambda}]\succeq
\chi[\rho_{a|x}] \quad\forall~a,x,\label{Eq:SRConstr1-Relaxed}\\
&\quad\,\,\,\,\chi[\sigma_{\lambda}]\succeq 0\quad \forall ~\lambda,\label{Eq:SRConstr2-Relaxed}\\
&\sum_a\chi[\rho_{a|x}] = \sum_a\chi[\rho_{a|x'}] ~~\forall\, x\neq x',\label{Eq:AssemblageRelaxed2}
\end{align}
\begin{align}
&\!\!\!\!\!\!\!\!\!\!\!\!\!\!\!\!\!\sum_a \chi[\rho_{a|x}]_\text{tr} = 1 ~~\forall ~x,\quad \chi[\rho_{a|x}]\succeq 0 ~~\forall ~a,x\label{Eq:AssemblageRelaxed3},\\
&\!\!\!\!\!\!\!\!\!\!\!\!\!P(a,b|x,y)=P_{\mbox{\tiny obs}}(a,b|x,y)\quad\forall\quad a,b,x,y.\label{Eq:CompatibleObservation}
\end{align}  
\end{subequations}
gives a lower bound on $\text{SR}(\{\rho_{a|x}\}_{a,x})$, and hence a lower bound on $\text{SR}(\rab)\ge\text{SR}(\{\rho_{a|x}\}_{a,x})$. In the above equation, Eqs.~\eqref{Eq:SRObj-Relaxed}-\eqref{Eq:SRConstr2-Relaxed} inherit directly from  Eqs.~\eqref{Eq:SRObj}-\eqref{Eq:SRConstr2} by applying the single-system mapping $\chi[\rho]=\Lambda(\rho)$, while Eqs.~\eqref{Eq:AssemblageRelaxed2} and \eqref{Eq:AssemblageRelaxed3} arise from physical constraints (consistent reduced states, normalization \& positivity) that have to be satisfied by the the assemblage $\SAB$. Note that a lower bound on SR$(\rho)$ can already be obtained if instead of Eq.~\eqref{Eq:CompatibleObservation}, we use some partial, but nontrivial information from $\vec{P}_{\mbox{\tiny obs}}$, such as the amount of a certain Bell inequality-violation, cf. Eq.~(\ref{relax_SR2}). Consider, for instance, the \revision{\em Clauser-Horne-Shimony-Holt} (CHSH) Bell inequality~\cite{Clauser69}:
\begin{equation}
	\mathcal{I}_{\mbox{\tiny CHSH}}: \SCHSH=E_{11}+E_{12}+E_{21}-E_{22}\stackrel{\mathcal{L}}{\leq} 2,	
\end{equation}
where $E_{xy}=\sum_{a,\revision{b}=1,2} (-1)^{a+b} P(a,b|x,y)$ is the two-partite correlator and $\mathcal{L}$ signifies that the bound holds for Bell-local correlations. By solving Eq.~(\ref{relax_SR2}) \revision{for} $\SCHSH=t>2$, our computation suggests the tight lower bound of 
\begin{equation}\label{Eq:SR:CHSH}
	\text{SR}(\rho|\SCHSH=t)\ge \frac{t-2}{2}\left(\sqrt{2}-1\right). 
\end{equation}
 Other explicit examples can be found in Appendix~\ref{App:SR}. Since SR$(\rho)$ provides a lower bound~\cite{Piani15} on the generalized robustness of entanglement~\cite{Vidal99,Steiner03} $R_g(\rho)$, the SDP of Eq.~\eqref{relax_SR} also provides a device-independent lower bound on $R_g(\rho)$ (cf.~\cite{Moroder13} for a lower bound based instead on negativity). Analogous relaxation also allows us to obtain device-independent lower bound on steerable weight; see Appendix~\ref{App:SW}.
  
{\em Device-independent quantification of measurement incompatibility.---}The device-independent estimate of SR$(\{{\rho}_{a|x}\}_{a,x})$ obtained above also gives a device-independent lower bound on the incompatibility of the measurements assemblage $\MAA$, as quantified using the \revision{\em incompatibility robustness} (IR)~\cite{Uola15}. In connection with the SDP formulation of the AMM, this will provide an estimate of the incompatibility of two or more measurements based on the violation of a Bell inequality, generalizing the quantitative relation found for the CHSH scenario by Wolf {\em et al.}~\cite{Wolf09}. To this end, we shall now prove that ${\rm IR}(\MAA) \geq  {\rm SR}(\SAB)$. Recall from~\cite{Uola15} that for any given measurement assemblage  $\GMA$, its IR is the minimum $t$ such that $\{ (M_{a|x} + t N_{a|x})/(1+t) \}_{x,a}$ is jointly measurable for some measurement assemblage $\{ N_{a|x} \}_{x,a}$. To see how one can relate SR$(\SAB)$ to IR$(\MAA)$, let $t_0=\text{IR}(\MAA)$, then by definition, there exists another measurement assemblage $\{N_{a|x}\}_{x,a}$ such that $\{(A_{a|x} + t_0 N_{a|x})/(1+t_0)\}_{x,a}$ is jointly measurable. Define ${\tau}_{a|x}=  \tr_\text{A}[(N_{a|x}\otimes \openone) \rab]$, then it is easy to see that ${\{(\rho_{a|x} + t_0 {\tau}_{a|x})/(1+t_0)\}_{a,x}}$ is an unsteerable assemblage --- the corresponding LHS $\{{\sigma}_\lambda\}_\lambda$ can be constructed from the joint POVM $\{G_\lambda\}_\lambda$ as ${\sigma}_{\lambda} =   \tr_\text{A}[(G_\lambda\otimes \openone) \rab]$. Since ${\rm SR}(\SAB)$ is defined as the smallest $t$ such that the mixture of $\SAB$ with some other state assemblage become unsteerable, we thus obtain 
\begin{equation}\label{Eq:IR-SR}
	{\rm IR}(\MAA) \geq  {\rm SR}(\SAB). 
\end{equation}

{\em Self-testing of measurements.---}By Eq.~\eqref{Eq:IR-SR} established above, the device-independent bounds obtained by solving the SDP of Eq.~\eqref{relax_SR} or Eq.~(\ref{relax_SR2}) provides us a handle to self-test measurements performed in a device-independent setting. For instance, for an observed CHSH-Bell violation of $2\sqrt{2}$, Eq.~\eqref{Eq:SR:CHSH} gives $(\sqrt{2}-1)^2$, which coincides precisely with the IR of a pair of mutually unbiased qubit measurements (see Appendix~\ref{App:MUB}). For a more complicated example, consider the ``elegant" Bell inequality which involves four (three) measurement settings on Alice's (Bob's) system~\revision{\cite{Christensen15,Gisin:ManyQuestions}}:
\begin{align}
\mathcal{I}_E:\mathcal{S}_{E} = &E_{11}+E_{12}+E_{13}+E_{21}-E_{22}-E_{23} \\ \nonumber  
						-&E_{31}+E_{32}-E_{33}-E_{41}-E_{42}+E_{43} \stackrel{\mathcal{L}}{\leq} 6.
\end{align}
Solving Eq.~(B5) assuming the maximal quantum violation of $\mathcal{I}_E$, i.e., $\mathcal{S}_E=4\sqrt{3}$ gives a device-independent lower bound of SR$(\SAB)$, and hence IR$(\MAA)$ of 0.2679. On the other hand, if we solve the SDP by considering, instead, Alice's state assemblage, then one obtains a device-independent lower bound of SR$(\SAA)$, and hence IR$(\MAB)$ of 0.2440. These values  coincide (within numerical precision) with, correspondingly, the IR of qubit measurements that span a symmetrical tetrahedron and the IR of three mutually unbiased qubit measurements, as one expects from the optimal qubit measurements leading to the maximal quantum violation of $\mathcal{I}_E$. 
For a higher-dimensional example, see Appendix~\ref{App:SR}.

{\em  Discussion.---}While our method of assemblage moment matrices (AMM) has allowed us to estimate quantities of interest in an EPR-steering scenario, it is worth noting that---in contrast to the approach of~\cite{Kogias15,Law2014}---we do not impose any additional constraints on the moments based on knowledge of the (measurement) assemblage --- such information is {\em  unavailable} in a device-independent setting. Rather, we exploit the fact that any measurement assemblage necessarily induces a state assemblage on the remaining party \revision{ (parties)}, and the moment matrix of these {\em subnormalized} states contains information also about the correlation between their measurement outcomes. 

In the bipartite setting, given that the AMM only involve higher level moments for one of the parties, one may wonder whether the superset characterization based on AMM would converge to the set of quantum correlations. While this may seem extremely unlikely at first sight, our numerical results (see Appendix~\ref{App:TsirelsonBounds}) suggest that this may be plausible after all (see also~\cite{Doherty04} where it is sufficient to consider  one-party-extension in a bipartite problem). Answering this question clearly sheds light on the minimal set of moments needed to characterize the set of quantum distributions. Evidently, the AMM approach can also be generalized to the multipartite scenario and it will be interesting to see what quantities of interest it would allow us to estimate in a device-independent setting. In addition, by combining the approach of AMM and the recent works on moment matrices with dimension constraints~\cite{Navascues14,Navascues15}, it seems conceivable that one would obtain a set of tools that are naturally suited for the recently introduced concept of dimension-bounded steering~\cite{Moroder16}.

In this work, we have focused on the device-independent estimation of steerability based on steering robustness (SR). Apart from allowing us to self-test the incompatibility of the employed local measurements, SR has also provided an estimate---in a device-independent setting---the generalized robustness of entanglement and the usefulness of the underlying quantum state for a subchannel discrimination problem. Beyond the familiar example of quantum key distribution and randomness expansion, this is the first instance where a device-independent {\em source} characterization immediately allows us to quantify its usefulness for a quantum information processing task. Clearly, this will motivate further work in  connecting the various measures of nonlocality to their operational meaning(s) in quantum information processing. On the other hand, since all pure bipartite entangled states give maximal steerable weight~\cite{Piani15}, it seems plausible that a device-independent estimation of steerable weight---achievable using AMM---may allow one to self-test the purity of the underlying quantum state---a possibility that deserves further investigation.

Coming back to the self-testing of measurements, our examples have illustrated that our inequality relating SR and incompatibility robustness (IR), Eq.~\eqref{Eq:IR-SR}, can indeed be saturated---even in a device-independent setting---for certain assemblage arising from \revision{the} measurements on a maximally entangled state. What about a nonmaximally entangled state? 
Our preliminary investigation---using the partially entangled two-qutrit state~\cite{Acin2002} that maximally violates the $I_{2233}$ Bell inequality~\cite{Collins04}---suggests that while our device-independent estimate of SR using Eq.~\eqref{relax_SR} may be tight, inequality~\eqref{Eq:IR-SR} may be {\em strict} for the same assemblage, thereby leaving a gap in our device-independent estimate of IR using Eq.~\eqref{relax_SR}. The question of when Eq.~\eqref{Eq:IR-SR} is saturated therefore deserves a more thorough investigation in the future (see also Appendix~\ref{App:IntermediateIneq}). 
Finally, let us point that while the IR does provide---as our examples illustrate---a handy way to self-test incompatible measurements, it is not clear how to make the self-testing robust in terms of this figure of merit. In particular, it would be desirable to translate any finite deviation from the ideal value of IR in terms of other more familiar figures of merit, such as the fidelity, or say, the trace distance~\cite{Nielsen2011} with respect to the corresponding Choi-Jamio{\l}kowski state~\cite{Jamiokowski74,Choi75}.

\begin{acknowledgments}
The authors acknowledge useful discussions with Otfried G\"uhne, Matthew F. Pusey, Denis Rosset, Roope Uola and are grateful to Daniel Cavalcanti for helpful comments on an earlier version of this manuscript. This work is supported by the Ministry of Education, Taiwan, R.O.C., through ``Aiming for the Top University Project" granted to the National Cheng Kung University (NCKU), and the Ministry of Science and Technology, Taiwan (Grants No. 104-2112-M-006-021-MY3 and No. 103-2112-M-006-017-MY4), the EU (Marie Curie CIG 293993/ENFOQI), the FQXi Fund (Silicon Valley Community Foundation), and the DFG.

{\em Note added.---} While completing this manuscript, we became aware of the work of~\cite{Cavalcanti:1601.07450} which independently proved a connection between the incompatibility robustness of the untrusted party's POVM and a strengthened version of the steering robustness of the trusted party's assemblage, thereby providing an alternative way to estimate the steerability and hence measurement incompatibility using nonlocal correlations.
\end{acknowledgments}

\appendix

\section{More details on the assemblage moment matrix}
\label{App:AMM}

Consider a steering experiment in which Alice's measurement assemblage and the  state assemblage induced on Bob's side are, respectively, $\MAA$ and $\SAB$. Moreover, if Bob performs measurement $\MAB$ on this assemblage, then the completely positive map $\Lambda$ discussed in the main text reads as:
\begin{equation}\label{Eq:L1Mapping}
	\chi[\rho_{a|x}] = \sum_{ij}|i\rangle\langle j|\text{tr}[\rho_{a|x} B_j^\dagger B_i],
\end{equation}
where $B_i\in\id\bigcup\MAB\setminus\{B_{|\B| |y}\}_y$. For the purpose of device-independent characterization, by appealing to Neumark's extension, elements in the measurement assemblage $\MAB$ can be taken, without loss of generality, as projectors (see also Appendix \ref{App:Neumark}). This choice simplifies the associated SDPs (see Appendix~\ref{App:SDP}) by reducing the number of variables $\{u_v\}$ using the linear constraints $B_{b|y}B_{b'|y}= \delta_{b,b'} B_{b|y}$ (orthogonality and idempotence of projectors).

As discussed in details in~\cite{NPA2008}, for the characterization of the set of quantum distributions $P(a,b|x,y)$ based on moments, instead of considering operators $B_i\in\id\bigcup\MAB\setminus\{B_{|\B| |y}\}_y$, one may equivalently consider the set of operators $B_i\in\bigcup\MAB$. For this latter choice, we see that
\begin{align}
	\tr\{\chi[\rho_{a|x}]\} &= \sum_{i}\text{tr}[\rho_{a|x} B_i^\dagger B_i]
	=\sum_{b,y} \text{tr}[\rho_{a|x} B_{b|y}^\dagger B_{b|y}],\nonumber\\
	&=\sum_{b,y} \text{tr}[\rho_{a|x} B_{b|y}]=\tr[\rho_{a|x} ]\revision{|\mathcal{Y}|}.
\end{align}
Thus, the completely positive map discussed in the main text (as well as that introduced in~\cite{Moroder13}) can be made trace-preserving by considering the equivalent list of operators discussed above, and by absorbing the constant factor \revision{$|\mathcal{Y}|$} into the definition of the Kraus operators.

\subsection{An example of the $\ell=1$ assemblage moment matrices}
\label{App:AMM:Level1}
For the specific case where $|\mathcal{Y}|=|\B|=2$, and where $\{B_i\}_i=\{\id,B_{1|1},B_{1|2}\}$, Eq.~\eqref{Eq:L1Mapping} gives rise to the following matrix representation of $\chi[\rho_{a|x}]$ for each $a$ and $x$: 
\begin{equation}\label{Eq:MMatrix:level1}
\begin{split}
\begin{pmatrix}
\tr(\rho_{a|x}) & \tr(\rho_{a|x} B_{1|1}) & \tr(\rho_{a|x} B_{1|2})\\
\tr(\rho_{a|x} B_{1|1}) & \tr(\rho_{a|x} B_{1|1}) & \tr(\rho_{a|x}B_{1|1}^\dag B_{1|2})\\
\tr(\rho_{a|x} B_{1|2}) & \tr(\rho_{a|x} B_{1|2}^\dag B_{1|1})  & \tr(\rho_{a|x} B_{1|2})
\end{pmatrix},\\
=\begin{pmatrix}
\braket{A_{a|x}\otimes\id} & \braket{A_{a|x}\otimes B_{1|1}} & \braket{A_{a|x}\otimes B_{1|2}}\\
\braket{A_{a|x}\otimes B_{1|1}} & \braket{A_{a|x}\otimes B_{1|1}} & \braket{A_{a|x}\otimes B_{1|1}^\dag B_{1|2}}\\
\braket{A_{a|x}\otimes B_{1|2}} & \braket{A_{a|x}\otimes B_{1|2}^\dag B_{1|1}}  & \braket{A_{a|x}\otimes B_{1|2}}
\end{pmatrix}
\end{split}
\end{equation}
where the expectation value $\braket{O}$ for the operator $O$ is understood to be evaluated with respect to the shared quantum state $\rab$.

In the device-independent setting, neither the state $\rab$, nor the measurement assemblages $\MAA$, $\MAB$ are known. Nonetheless, some of these entries can be estimated from the observed correlations between measurement outcomes, $P(a,b|x,y)$. In this notation  we may thus rewrite Eq.~\eqref{Eq:MMatrix:level1} as:
\begin{equation}\label{Eq:MMatrix:level1:DI}
\chi[\rho_{a|x}]=
\begin{pmatrix}
P(a|x) & P(a,1|x,1) & P(a,1|x,2)\\
P(a,1|x,1) & P(a,1|x,{\blu 1}) & u_1+iu_2\\
P(a,1|x,2) & u_1-iu_2  & P(a,1|x,2)
\end{pmatrix},
\end{equation}
where we have expressed the experimentally inaccessible expectation value as:
\begin{equation}
\braket{A_{a|x}\otimes B_{1|1}^\dag B_{1|2}}=\tr(\rab A_{a|x}\otimes B_{1|1}^\dag B_{1|2})=u_1+iu_2,
\end{equation}
with $u_v$ \revision{being} a real number.

\section{SDPs arising from the assemblage moment matrix approach}
\label{App:SDP}
By considering the necessary conditions for a given collection of matrices to represent the assemblages moment matrices, one obtains a hierarchy of {\em superset} characterization of the set of quantum $P(a,b|x,y)$, as with the NPA moment matrices~\cite{NPA,NPA2008}.

\subsection{Existence of a quantum representation for $P_{\mbox{\tiny obs}}(a,b|x,y)$}

Due to the complete-positivity of the mapping $\Lambda$, for any given level $\ell$, the resulting moment matrix $\chi[\rho_{a|x}]$ must be positive semidefinite. Thus, for any legitimate quantum distribution $P(a,b|x,y)$, there must exist choices of unobservable expectation values (e.g., those expressed in terms of $u_1$ and $u_2$ in the example given in Appendix~\ref{App:AMM:Level1}) such that each of the assemblage moment matrices is positive semidefinite. 
Thus, a necessary condition for an observed distribution $P_{\mbox{\tiny obs}}(a,b|x,y)$ to admit a quantum representation 
\begin{equation}
	P_{\mbox{\tiny obs}}(a,b|x,y) =\tr(\rab A_{a|x}\otimes B_{b|y}), 
\end{equation}
is to admit a feasible solution to the SDP feasibility problem
\begin{equation}\label{SDP:QMP}
\begin{aligned}
\text{Find} ~~& \{u_v\} \\
\text{such that}~~ & \chi[\rho_{a|x}]\geq 0 ~~\forall ~a,x\\
&\sum_a\chi[\rho_{a|x}] = \sum_a\chi[\rho_{a|x'}] ~~\forall x\neq x'\\
&\sum_a \chi[\rho_{a|x}]_\text{tr} = 1 ~~\forall ~x\\
&P(a,b|x,y)=P_{\mbox{\tiny obs}}(a,b|x,y)\quad\forall\quad a,b,x,y.
\end{aligned}  
\end{equation}
In other words, for any given level $\ell$, if there does not exist any choice of real variables $\{u_v\}$ such that the above SDP is feasible, then the observed correlation $P_{\mbox{\tiny obs}}(a,b|x,y)$ is certifiably non-quantum.

Notice that for any feasible solution $\{u_v\}$ to the above SDP, if the subset of $\{u_v\}$ that are associated with the imaginary components of $\chi[\rho_{a|x}]$ are all negated, the resulting moment matrix becomes $\chi[\rho_{a|x}]^{\mbox{\tiny T}}$, i.e., the transpose of the original matrix, which also represents a feasible solution to the above SDP. As a result, their average $\frac{1}{2}\left(\chi[\rho_{a|x}]+\chi[\rho_{a|x}]^{\mbox{\tiny T}}\right)$ must also be a feasible solution to the above SDP. Hence, without loss of generality, it is sufficient to consider real assemblage moment matrices in \revision{the} SDP of Eq.~\eqref{SDP:QMP} as well as all the SDPs discussed below.

\subsection{Upper bounds on Tsirelson bound}

Using the assemblage moment matrices of any given level, an upper bound on the Tsirelson bound of a Bell inequality:
\begin{equation}\label{Eq:BI}
	\beta^{xy}_{ab} P(a,b|x,y)\stackrel{\mathcal{L}}{\le} L
\end{equation}
can be calculated by solving the following SDP:
\begin{equation}\label{SDP:Tsirelson}
\begin{aligned}
\max_{P(a,b|x,y),u_v} ~~& \sum_{x,y,a,b} \beta^{xy}_{ab} P(a,b|x,y) \\
\text{subject to }~~ & \chi[\rho_{a|x}]\geq 0 ~~&&\forall ~a,x\\
&\sum_a\chi[\rho_{a|x}] = \sum_a\chi[\rho_{a|x'}] ~~&&\forall x\neq x'\\
&\sum_a \chi[\rho_{a|x}]_\text{tr} = 1 ~~&&\forall ~x\\
\end{aligned}  
\end{equation}

\subsection{Lower bounds on the steering robustness for a given Bell violation}

As mentioned in the main text, instead of assuming full knowledge of the quantum distribution $P(a,b|x,y)$ and solving the SDP given in Eq.~(5), non-trivial lower bound on the steerability can already be obtained by solving the following SDP:
\begin{subequations}\label{relax_SR2}
\begin{align}
\min_{\{u_v\}} ~~&\left(\sum_{\lambda}\chi[\sigma_{\lambda}]_\text{tr}\right)-1 \\
\text{subject to }~~ &\sum_{\lambda}D(a|x,{\lambda})\chi[\sigma_{\lambda}]\succeq
\chi[\rho_{a|x}] \,\,\forall~a,x,\\
&\quad\,\,\,\,\chi[\sigma_{\lambda}]\succeq 0\quad \forall ~\lambda,\\
&\sum_a\chi[\rho_{a|x}] = \sum_a\chi[\rho_{a|x'}] ~~\forall\, x\neq x',\\
&\!\!\!\!\!\!\!\!\!\!\!\!\!\!\!\!\!\!\!\!\sum_a \chi[\rho_{a|x}]_\text{tr} = 1 ~~\forall ~x,\quad \chi[\rho_{a|x}]\succeq 0 ~~\forall ~a,x,\\
&\quad \sum_{x,y,a,b} \beta^{xy}_{ab} P(a,b|x,y) = \mathcal{S}_{\mbox{\tiny obs}}\label{Eq:SObs},
\end{align}  
\end{subequations}
based on \revision{the} observed quantum value $\mathcal{S}_{\mbox{\tiny obs}}$ of certain Bell inequality, cf. Eq.~\eqref{Eq:BI}.

\subsection{Lower bounds on the steerable weight}
\label{App:SW}

In analogy to the SDP presented in Eq.~(5) for obtaining a device-independent lower bound on the steering robustness, a device-independent lower bound on the steerable weight~\cite{SNC14} of the underlying state can be obtained by solving the following SDP:
\begin{subequations}\label{SDP:SW}
\begin{align}
\min_{\{u_v\}} ~~&1-\left(\sum_{\lambda}\chi[\sigma_{\lambda}]_\text{tr}\right) \label{Eq:SWObj-Relaxed}\\
\text{subject to }~~ &\chi[\rho_{a|x}]\succeq \sum_{\lambda}D(a|x,{\lambda})\chi[\sigma_{\lambda}]
 \,\,\forall~a,x,\label{Eq:SRConstr1-Relaxed}\\
&\quad\,\,\,\,\chi[\sigma_{\lambda}]\succeq 0\quad \forall ~\lambda,\label{Eq:SWConstr2-Relaxed}\\
&\sum_a\chi[\rho_{a|x}] = \sum_a\chi[\rho_{a|x'}] ~~\forall\, x\neq x',\\
&\!\!\!\!\!\!\!\!\!\!\!\!\!\!\!\!\!\!\!\sum_a \chi[\rho_{a|x}]_\text{tr} = 1 ~~\forall ~x,\quad \chi[\rho_{a|x}]\succeq 0 ~~\forall ~a,x,\\%
&\!\!\!\!\!\!\!\!\!\!\!\!\!P(a,b|x,y)=P_{\mbox{\tiny obs}}(a,b|x,y)\quad\forall\quad a,b,x,y.\label{Eq:CompatibleObservation:SW}
\end{align}  
\end{subequations}
Similarly, as with the estimation of steering robustness, a lower bound on the steerable weight can already be obtained if the constraint of Eq.~\eqref{Eq:CompatibleObservation:SW} is replaced by Eq.~\eqref{Eq:SObs}.

\section{Computation results}

\subsection{Upper bound(s) on Tsirelson bounds}
\label{App:TsirelsonBounds}

We consider the CHSH~\cite{Clauser69} inequality, the $I_{3322}$ inequality~\cite{Collins04}, the $I_{2233}$ inequality~\cite{Collins04}, and the set of known facet-defining Bell inequalities defined for the $\{[3\,3\,3]\,[3\, 3\, 3]\}$ Bell scenario~\cite{Schwarz16}. The results are shown in Tables~\ref{TB2} and \ref{TB1}. In most of these cases, the upper bounds obtained by solving Eq.~\eqref{SDP:Tsirelson} already converge to the Tsirelson bound~\cite{Tsirelson1980} at the 2nd level of AMM. \revision{The only exceptions to these are (1) the $I_{3322}$ inequality where there remains a gap of about $6\times10^{-6}$ between our SDP bound at level $\ell=4$ and the (tight) SDP upper bound known in the literature~\cite{Pal10} and (2) inequality 16, 17 and 19 from~\cite{Schwarz16} where there remain, in the first two cases a difference of the order of $10^{-3}$ and in the last case a difference of the order of $10^{-2}$ between our $\ell=2$ SDP bound and the (tight) SDP upper bound known in the literature.}

\begin{table}[h!]
\centering
\caption{Upper bounds on the Tsirelson bounds computed using Eq.~\eqref{SDP:Tsirelson} for the facet-defining Bell inequalities listed in~\cite{Schwarz16}. These 19 inequalities are defined for the $\{[3\,3\,3]\,[3\,3\,3]\}$ scenario, i.e., when both  parties can each employ three ternary-outcome measurements [see Eq.~(6) and TABLE I in~\cite{Schwarz16}]. The second column below ($\mathcal{S}_\text{max}^\mathcal{Q}$, i.e., Tsirelson's bound) is extracted from the 4th column of TABLE II in~\cite{Schwarz16} \revision{whereas the 3rd-5th column gives, respectively, our computational result for $\ell=3,2$ and $1$.}  
$^\dag$\revision{For $n=16$ and $19$,} the best upper bound that we obtained from Eq.~\eqref{SDP:Tsirelson} does not match with the best quantum violation known. \revision{Entries marked with ``-" are those where the SDP bound for that level was not computed as a lower-level relaxation already gave a tight bound.}
}
\begin{tabular}{|c|c|c|c|c|}
\hline
$n$ & $\mathcal{S}_\text{max}^\mathcal{Q}$ & \revision{$\ell=3$} & $\ell=2$ & $\ell=1$  \\ \hline \hline
$1$ & 2.6972 & - & 2.6972 & 2.7303  \\ \hline
$2$ & 2.6712 & - & 2.6712 & 2.7024  \\ \hline
$3$ & 2.6586  & -  & 2.6586 & 2.7732  \\ \hline
$4$ & 2.6586 & -  & 2.6586 & 2.7730  \\ \hline
$5$ & 2.6488 & -  & 2.6488 & 2.7453  \\ \hline
$6$ & 2.6577 & - & 2.6577 & 2.7253  \\ \hline
$7$ & 2.6577 & - & 2.6577 & 2.7470  \\ \hline
$8$ & 2.6577 & - & 2.6577 & 2.7255  \\ \hline
$9$ & 2.6720 & - & 2.6720 & 2.7574  \\ \hline
$10$& 1.6720 & - & 1.6720 & 1.7352  \\ \hline
$11$& 2.6955 & - & 2.6955 & 2.7392  \\ \hline
$12$& 2.5820 & - & 2.5820 & 2.6120  \\ \hline
$13$& 2.6712 & - & 2.6712 & 2.7141  \\ \hline
$14$& 2.6972 & - & 2.6972  & 2.7382  \\ \hline
$15$& 1.5923 & -  & 1.5923 & 1.6030  \\ \hline
$16$& 1.2532 & 1.2534$^\dag$ & 1.2609 & 1.3183  \\ \hline
$17$& 1.3090 & 1.3090 & 1.3162 & 1.4294  \\ \hline
$18$& 1.4142 & - & 1.4142 & 1.4374  \\ \hline
$19$& 1.3782 & 1.3899$^\dag$ & 1.4092 & 1.5837  \\ \hline
\end{tabular}\label{TB2}
\end{table}

\begin{table}[h!]
\centering
\caption{Upper bounds on the Tsirelson bounds for the CHSH-Bell inequality, the $I_{3322}$ inequality, and the $I_{2233}$ inequality. \revision{$^\dag$For the $I_{3322}$ inequality, there remains a non-negligible gap between our bound and the best SDP bound known in the literature~\cite{Pal10}. Entries marked with ``-" are those where the SDP bound for that level was not computed as a lower-level relaxation already gave a tight bound.}}
\begin{tabular}{|l|l|c|c|c|c|cl}
\hline
Bell inequality & Tsirelson bound  & $\ell=4$ & $\ell=3$ & $\ell=2$ & $\ell=1$ \\ \hline \hline
CHSH    & 2.8284 & -   & -  & -  & 2.8284 \\ \hline
$I_{3322}$ & 0.2509~\cite{Pal10}   & 0.2509$^\dag$ & 0.2512 & 0.2550   & 0.3621  \\ \hline
$I_{2233}$ & 0.3050 & -  & -  & 0.3050   & 0.3078 \\ \hline
\end{tabular}\label{TB1}
\end{table}

\begin{table}[h!]
\centering
\caption{\revision{Size of the SDP for solving an upper bound on the Tsirelson sbound at level $\ell_{\mbox{\tiny M}}$ of the hierarchy defined by Moroder~{\em et al.}~\cite{Moroder13}. The leftmost column gives the Bell scenario using the notation defined in~\cite{Barnea2013}. The size of each SDP is specified in terms of a pair $(D, N)$ of numbers with $D$ being the dimension of the SDP matrix, and $N$ being the number of independent moments. Entries marked with ``-" are those where we do not have any information about the size of the particular SDP.}}
\begin{tabular}{|c|c|c|c|c|c|c|cl}
\hline
Scenarios  & $\ell_{\mbox{\tiny M}}=3$ & $\ell_{\mbox{\tiny M}}=2$ & $\ell_{\mbox{\tiny M}}=1$ \\ \hline \hline
$\{[2\, 2]\,\,[2\, 2]\}$   & (49, 108)   & (25, 52)  & (9, 16)  \\ \hline
$\{[2\,2\,2]\,\,[2\, 2\,2]\}$    & (484, 18\,291)   & (100, 1107)  & (16, 57)   \\ \hline
$\{[3\, 3]\,\,[3\, 3]\}$    & (841, 32\,424)   & (169, 1944)  & (25, 96)  \\ \hline
$\{[3\,3\,3]\,\,[3\, 3\,3]\}$   & -   & (961, 131\,040)  & (49, 504)   \\ \hline
\end{tabular}\label{Complexity:Moroder}
\end{table}

\begin{table}[h!]
\centering
\caption{\revision{Size of the SDP described in Eq.~\eqref{SDP:Tsirelson} at various level $\ell$ of our hierarchy. The leftmost column gives the Bell scenario using the notation defined in~\cite{Barnea2013} while the second column gives the number of blocks $N_{\mbox{\tiny blk}}$ involved in the SDP. The size of each SDP is specified in terms of a pair $(D_{\mbox{\tiny blk}}, N)$ of numbers with $D_{\mbox{\tiny blk}}$ being the dimension of each AMM block (i.e., each positive-semidefinite block), and $N$ being the total number of distinct moments before taking into account of the equality constraints given in Eq.~\eqref{SDP:Tsirelson}. The full (block-diagonal) SDP matrix thus has dimension of $D_{\mbox{\tiny blk}} N_{\mbox{\tiny blk}}$.} }
\begin{tabular}{|c|c|c|c|c|c|c|c|cl}
\hline
Scenarios & $N_{\mbox{\tiny blk}}$ & $\ell=4$ & $\ell=3$ & $\ell=2$ & $\ell=1$ \\ \hline \hline
$\{[2\, 2]\,\,[2\, 2]\}$   & 4& (81, 48)   & (27, 36)  & (9, 24) &  (3, 12)  \\ \hline
$\{[2\,2\,2]\,\,[2\, 2\,2]\}$ & 6   & (256, 2427)   & (64, 627)  & (16, 159) & (4, 33)   \\ \hline
$\{[3\, 3]\,\,[3\, 3]\}$    & 6 & (625, 3240)   & (125, 840)  & (25, 216)  & (5, 48)\\ \hline
$\{[3\,3\,3]\,\,[3\, 3\,3]\}$  & 9 & -   & (549, 45747) & (49, 2439)  & (7, 171)   \\ \hline\hline
\end{tabular}\label{Complexity:AMM}
\end{table}

\revision{Let's now briefly comment on the convergence of these Tsirelson bounds based on the AMM formulation compared with that of the SDPs derived from the hierarchy proposed in~\cite{Moroder13}. Since higher moments are included only for Bob's measurements, the convergence to the quantum value, as one would expect, is generally slower than that compared with the Moroder's hierarchy (or with the NPA hierarchy~\cite{NPA2008}). For instance, with the $I_{2233}$ inequality, which is equivalent to the Collins-Gisin-Linden-Massar-Popescu inequality~\cite{CGLMP} (see~\cite{Liang:PhDthesis} for a proof), convergence is already achieved at $\ell_{\mbox{\tiny M}}=1$ (see~\cite{NPA2008}) but for the SDP of Eq.~\eqref{SDP:Tsirelson}, convergence to the quantum value seems to happen only with a considerably more complicated SDP (at $\ell=2$) --- see Table~\ref{Complexity:Moroder} and Table~\ref{Complexity:AMM} for the details of the size of these SDPs. Likewise for the $I_{3322}$ inequality, where the SDP bound at $\ell_{\mbox{\tiny M}}=2$ (giving 0.250\,875\,42) is already better than our considerably more complicated SDP bound at $\ell=4$ (giving 0.250\,881\,30).

} 

\revision{Finally, let us note} that, in contrast to the approach of~\cite{Teddei:1603.05247}, we do not assume any knowledge of the assemblage in computing the \revision{above} bounds.

\subsection{Steering robustness}
\label{App:SR}

Here we provide the steering robustness bounds that were computed based on the observed violation $\mathcal{S}_{\mbox{\tiny obs}}$ of an arbitrary, but given Bell inequality [cf Eq.~\eqref{Eq:BI}] via Eq.~\eqref{relax_SR2}. 
Our results for the CHSH-Bell inequality, the $I_{3322}$ inequality and the $I_{2233}$ inequality are summarized in the following figures.

\subsubsection{Minimal steering robustness certifiable by the observed Bell violation for three Bell inequalities}

\begin{figure}[h!]
\emph{\includegraphics[width=7.5cm]{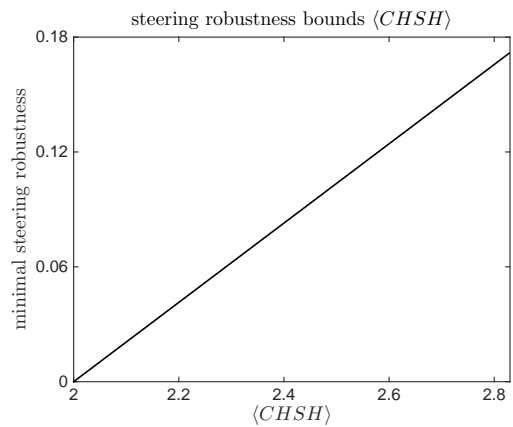} }
\caption{Steering robustness lower bound(s) for given violation of the CHSH-Bell inequality via AMM with  $\ell=1,2,3$.}
\label{SR_CHSH}
\end{figure}

\begin{figure}[h!]
\emph{\includegraphics[width=7.5cm]{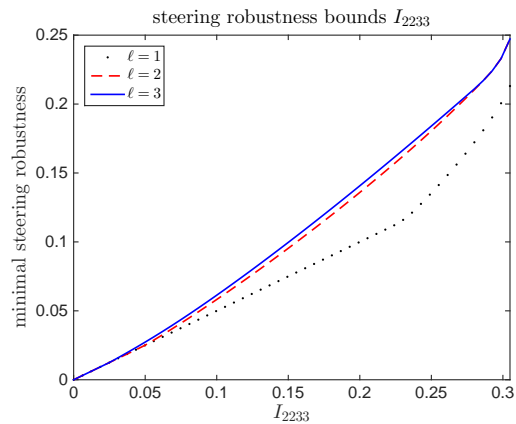} }
\caption{Steering robustness lower bound(s) for given violation of the $I_{2233}$ Bell inequality (involving two ternary-outcome measurements per party)  via AMM with  $\ell=1,2,3$.}
\label{SR_I2233}
\end{figure}

\begin{figure}[h!]
\emph{\includegraphics[width=7.5cm]{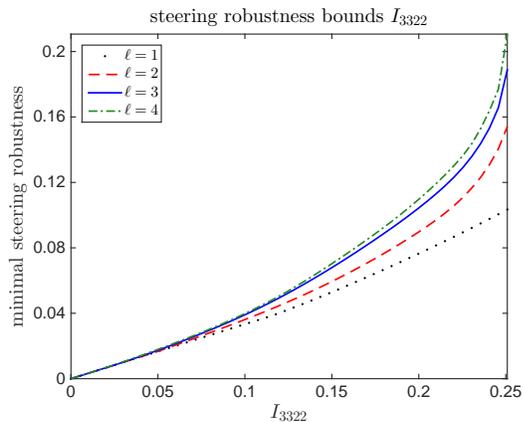} }
\caption{Steering robustness lower bound(s) for given violation of the $I_{3322}$ Bell inequality (involving three binary-outcome measurements per party) via AMM with  $\ell=1,2,3,4$.}
\label{SR_I3322}
\end{figure}

In reading \revision{these} figures, it is worth noting that if for any (intermediate) level, we obtain a straight line [cf Fig.~\ref{SR_CHSH}] joining two end points that each gives a tight lower bound (i.e., these lower bounds are indeed achievable in quantum theory), then the rest of the lower bounds are also tight and thus achievable in quantum theory. To see this, it is sufficient to note that the set of non-steerable quantum states (i.e., those admitting a LHS model) and hence the set of non-steerable state assemblages is convex. Thus, if the lower bound on the two ends points (corresponding, e.g., to the point with no quantum violation and the point with maximal quantum violation) are both quantum realizable, respectively, using quantum state $\rho_1$ with measurement assemblage $\GMA$ and quantum state $\rho_2$ with measurement assemblage $\{\widetilde{M}_{a|x}\}_{x,a}$, then the rest can also be realized by considering appropriate convex mixture of $\rho_1$ and $\rho_2$ in orthogonal subspace and the measurement $\{M_{a|x}\oplus\widetilde{M}_{a|x}\}_{x,a}$.

\subsubsection{Minimal steering robustness based on the maximal quantum violation of the $I^+_3$ Bell inequality~\cite{Liang09}}

Finally, we have also solved Eq.~\eqref{relax_SR2} by assuming the maximal quantum violation of a 3-setting, 3-outcome generalization of the CHSH-Bell inequality (termed $I^+_3$ in~\cite{Liang09}). The obtained device-independent lower bounds on steering robustness are summarized in the table below:
\begin{equation}\label{Eq:IdpConvergence}
	\begin{tabular}{|c|c|c|}\hline
	$\ell=2+$ & $\ell=2$ & $\ell=1$  \\ \hline \hline
	 0.4016 & 0.3070 & 0.1560 \\ \hline
	\end{tabular}
\end{equation}
The best lower bound presented above is obtained by considering AMM of size $243\times243$.

On the other hand, it is known that this maximal quantum violation is achievable by performing three mutually unbiased qutrit measurements on a pair of maximally entangled two qutrit state. The actual IR for these measurements as well as the SR of the corresponding state assemblage can be numerically determined by solving an SDP, both giving 0.4037.\footnote{Hence, inequality~(8) is also saturated in this case.} While our calculations do not quite close the gap between the actual value of IR for these measurements and the corresponding device-independent lower bounds, the rate of convergence of the above calculation, cf Eq.~\eqref{Eq:IdpConvergence}, makes it plausible that the SDP value for $\ell=3$ would coincide with the actual value within numerical precision.

\section{Projective measurements for Bob's assemblage}
\label{App:Neumark}

The use of projective measurements in the definition of the moment matrix has the advantage of simplifying the SDP by reducing the number of variables using additional linear constraints (orthogonality and idempotence of projectors). 


The Neumark dilation theorem (see, e.g.,~\cite{WolfLect}) allows one to extend any POVM to a projective measurement on a larger Hilbert space, and it has been used in connection with moment matrix approaches (cf.~\cite{NPA2008,Moroder13}). However, a naive approach to the problem has been shown to give rise to inconsistency both in moment matrix approaches \cite{Budroni13} and in relation with the joint measurability problem \cite{Heunen14}.

The critical point is to prove that for different POVMs there exists a joint Neumark dilation in the same Hilbert space. For the assemblage moment matrix, the proof is quite simple, and we include it here for completeness.

Consider the measurement assemblage for Bob $\{ B_{b|y}\}_{b,y}$ acting on $\mathbb{C}^d$. Each POVM $\{ B_{b|y}\}_b$ can be
refined to contain only rank-1 operators (e.g., by spectral decomposition), i.e.,  $B_{b|y}=\sum_{i\in I_{by}} \ket{\psi_i^y}\bra{\psi_i^y}$ and
$\sum_{i\in I_y} \ket{\psi_i^y}\bra{\psi_i^y} = \openone_d$, where $d$ is the dimension of Bob's system, $I_y = \cup_b I_{by}$, $I_{by}$ is defined by the above equation and $\ket{\psi_i^y}$ are generally subnormalized vectors. Possibly by adding some null vector,
we can assume that we use exactly $n$ vectors for any $\{ B_{b|y}\}_b$. The problem of finding a common Neumark dilation for the
assemblage  $\{ B_{b|y}\}_{b|y}$ is then equivalent to finding a common Neumark dilation for
$\{ \{\ket{\psi_i^y}\bra{\psi_i^y}\}_{i=1}^n\}_y$. Now, applying the Neumark dilation to each POVM
$\{\ket{\psi_i^y}\bra{\psi_i^y}\}_{i=1}^n$ (see, e.g.,~\cite{WolfLect}), we obtain  projective measurements $\{\ket{\phi_i^y}\bra{\phi_i^y}\}_{i=1}^n$ acting on the common space $\mathbb{C}^n$. The standard technique consists in taking a basis $\{\ket{j}\}_{j=1}^n$ of $\mathbb{C}^n$ and
defining the mapping $\Psi^y$ as $\Psi_{ij}^y=\mean{j | \psi_i^y}$, where  $\{\ket{j}\}_{j=1}^d$ is a basis of the embedding of $
\mathbb{C}^d$.  By construction, $\Psi^y$ is an isometry, i.e., $\Psi^{y\dagger}\Psi^y=\openone_d$, hence it can extended to a
unitary $\tilde{\Psi}^y$ acting on $\mathbb{C}^n$. The vectors $\ket{\phi_i^y}$ are then obtained as $\tilde{\Psi}^y_{i,j}=\mean{j | \phi_i^y}$, providing a Neumark dilation for the whole measurement assemblage $\{ B_{b|y}\}_{b,y}$ in the common Hilbert space.

\section{Inequalities bounding steering-equivalent observables}
\label{App:IntermediateIneq}

Here, we give an alternate proof of the connection between ${\rm IR}(\MAA)$ and ${\rm SR}(\SAB)$ which allows us to show that the incompatibility robustness of  Bob's steering-equivalent (SE) observables~\cite{Uola15} is actually sandwiched between these two quantities. Recall from~\cite{Uola15} that the steerability of the state assemblage $\SAB$  is equivalent to the {\em non}-joint-measurability of the corresponding SE observables for Bob, defined as
\begin{equation}\label{bobobs}
	\widetilde{B}_{a|x} = (\tilde{\rho}_{\mbox{\tiny B}})^{-\frac{1}{2}}\ \tilde{\rho}_{a|x}\ (\tilde{\rho}_{\mbox{\tiny B}})^{-\frac{1}{2}},
\end{equation}
where
\begin{equation}\label{Eq:tilderho}
\begin{split}
	&\,\,\tilde{\rho}_{a|x} = \Pi_{\mbox{\tiny B}} \rho_{a|x} \Pi_{\mbox{\tiny B}}^\dag,\\
	\tilde{\rho}_{\mbox{\tiny B}}&= \sum_a \tilde{\rho}_{a|x} =\Pi_{\mbox{\tiny B}} \rho_{\mbox{\tiny B}} \Pi_{\mbox{\tiny B}}^\dag,
\end{split}
\end{equation}
and ${\Pi_{\mbox{\tiny B}}: \mathcal{H}_{\mbox{\tiny B}} \rightarrow \mathcal{K}_{\rho_{\mbox{\tiny B}}}\subset \mathcal{H}_{\mbox{\tiny B}}}$ is the projection onto the subspace $\mathcal{K}_{\rho_{\mbox{\tiny B}}} := {\rm range}(\rho_{\mbox{\tiny B}})$ (cf.~\cite{Uola15}).

Let us first prove that ${\rm IR}(\{\widetilde{B}_{a|x}\}_{x,a})\leq {\rm IR}(\MAA)$.
Given the measurement assemblage $\MAA$, for any $t\ge0$ and measurement assemblage $\{N_{a|x}\}_{x,a}$ 
such that $\{(A_{a|x} + t N_{a|x})/(1+t)\}$ is jointly-measurable (JM), it can be shown that 
${\{(\widetilde{B}_{a|x} + t \widetilde{N}_{a|x})/(1+t)\}}$ is also JM for the choice of 
\begin{equation}\label{Eq:Ntilde_ax}
	{\widetilde{N}_{a|x}= (\tilde{\rho}_{\mbox{\tiny B}})^{-\frac{1}{2}}\Pi_{\mbox{\tiny B}} \tr_\revision{\text{A}}[(N_{a|x}\otimes \openone) \rab]\ \Pi_{\mbox{\tiny B}}^\dag (\tilde{\rho}_{\mbox{\tiny B}})^{-\frac{1}{2}}}.
\end{equation}
To see this, note that $\widetilde{N}_{a|x}$ is indeed a valid measurement assemblage since $\widetilde{N}_{a|x}\succeq0$ for all $x,a$ and $\sum_a \widetilde{N}_{a|x} = \revision{\openone_{\mathcal{K}_{\rho_{\mbox{\tiny B}}}}}$. Let us denote by $\{G_\lambda\}_\lambda$ the joint POVM of $\{(A_{a|x} + t N_{a|x})/(1+t)\}_{x,a}$, then from the defining equation of joint-measurability given in Eq.~(1), we get
\begin{equation}
	\frac{A_{a|x} + t N_{a|x}}{1+t} = \sum_\lambda D(a|x,\lambda) G_\lambda 
\end{equation}
It then follows from Eqs.~\eqref{bobobs}-\eqref{Eq:Ntilde_ax} that
\begin{equation}\label{Eq:BobSEObs:JM}
	\frac{\widetilde{B}_{a|x} + t \widetilde{N}_{a|x}}{1+t} = \sum_\lambda D(a|x,\lambda) \widetilde{G}_\lambda.
\end{equation}
where 
\begin{equation}
	{\widetilde{G}_\lambda = (\tilde{\rho}_\sB)^{-\frac{1}{2}}\Pi_\sB \tr_A[(G_\lambda \otimes \openone) \rab]\ \Pi_\sB^\dag (\tilde{\rho}_\sB)^{-\frac{1}{2}}}
\end{equation}
is the joint-POVM of the measurement assemblage ${\{(\widetilde{B}_{a|x} + t \widetilde{N}_{a|x})/(1+t)\}}_{x,a}$. Since IR$(\{\widetilde{B}_{a|x}\}_{x,a})$ is the smallest $t$ such that Eq.~\eqref{Eq:BobSEObs:JM} holds true (possibly for other choice of $\widetilde{N}_{a|x}$), we have ${\rm IR}(\{\widetilde{B}_{a|x}\}_{x,a})\leq {\rm IR}(\MAA)$. 

In an identical way, one can prove that ${\rm SR}(\SAB)\leq {\rm IR}(\{\widetilde{B}_{a|x}\}_{x,a})$. Firstly, we note that SR$(\SAB)={\rm SR}( \{\tilde{\rho}_{a|x}\}_{a,x}$ since $\rho_{a|x}$ vanishes on ${\rm range}(\rho_\sB)^\perp$.
Then, given the measurement assemblage $\{\widetilde{B}_{a|x}\}_{x,a}$, for any $t\ge0$ and measurement assemblage $\{O_{a|x}\}_{x,a}$
such that $\{(\widetilde{B}_{a|x} + t O_{a|x})/(1+t)\}_{x,a}$ is JM, it can be shown that 
$\{(\tilde{\rho}_{a|x} + t \tilde{\tau}_{a|x})/(1+t)\}$ admits a LHS model [cf Eq.~(2)] if we take $\tilde{\tau}_{a|x}= (\tilde{\rho}_\sB)^{\frac{1}{2}}\ O_{a|x}\ (\tilde{\rho}_\sB)^{\frac{1}{2}}$. Notice that $\tilde{\tau}_{a|x}$ is a valid assemblage because $\sum_a \tilde{\tau}_{a|x} = \tilde{\rho}_\sB$ and the corresponding LHS $\{\tilde{\sigma}_\lambda\}_\lambda$ can be constructed from the joint-POVM
$\{G_\lambda\}_\lambda$ as $\tilde{\sigma}_{\lambda} = (\tilde{\rho}_\sB)^{\frac{1}{2}}\  G_\lambda\ (\tilde{\rho}_\sB)^{\frac{1}{2}}$. Steps in the proof proceed analogously as those in the proof of ${\rm IR}(\{\widetilde{B}_{a|x}\}_{a,x})\leq {\rm IR}(\{A_{a|x}\}_{a,x})$.

The argument does not work in the other direction, because, in general, $(\tilde{\rho}_\sB)^{-\frac{1}{2}}\ \tilde{\tau}_{a|x}\ (\tilde{\rho}_\sB)^{-\frac{1}{2}}$, will not give a valid POVM, i.e., $\sum_{a} (\tilde{\rho}_\sB)^{-\frac{1}{2}}\ \tilde{\tau}_{a|x}\ (\tilde{\rho}_\sB)^{-\frac{1}{2}}\neq \openone$. 

All in all, we thus obtain the chain of inequalities:
\begin{equation}
\begin{split}
	{\rm IR}(\MAA) \geq {\rm IR}(\{\widetilde{B}_{a|x}\}_{x,a}) \geq  {\rm SR}(\SAB).
\end{split}
\end{equation}

\section{Analytical proof of IR for a pair of qubit measurements in  MUBs}
\label{App:MUB}

Given a measurement assemblage $\{A_{a|x}\}_{a=\pm1,x=1,2}$, of measurements in two mutually unbiased bases for a qubit, e.g., $A_{\pm|1}= (\openone \pm \sigma_x)/2$ and $A_{\pm|2}= (\openone \pm \sigma_z)/2$. We want to compute the minimal $t$ such that $\{O_{a|x} := (A_{a|x} + t N_{a|x})/(1+t)\}_{x,a}$ is jointly measurable. We can use the joint measurability criterion by Busch \cite{Busch86}, 
\begin{equation}\label{buschcond}
\| \vec r_1 +  \vec r_2\| + \|\vec r_1 - \vec r_2\| \leq 2 
\end{equation} 
where $r_1,r_2$ are the Bloch vector representation of the $+$ outcome for the observables $O_{a|x}$, $x=1,2$, namely $O_{+|x} = \frac{1}{2}\left[(1+\alpha_x)\openone+\vec r_x\cdot\vec\sigma\right]$. Let us consider for now the case of an unbiased observable, i.e., $\alpha_x=0$, obtained via mixing with  $N_{\pm|x} = \frac{1}{2}(\openone\pm\vec q_x\cdot\vec\sigma)$

Eq.~\eqref{buschcond} provides a necessary conditions for joint measurability \cite{Busch08}, which is also sufficient for the case of unbiased observables. We can then rewrite Eq.~\eqref{buschcond} as
\begin{equation}\label{busch2}
\frac{1}{1+t}\left(\ \| \vec x +  \vec z +t(\vec{q_1}+\vec{q_2}) \| + \|\vec x -  \vec z +t(\vec{q_1}-\vec{q_2}) \|\ \right)\leq 2, 
\end{equation}
where $\vec x, \vec z$ are the Bloch vectors associated with $A_{+|1,2}$. For fixed $t$, both terms on the l.h.s. of Eq.~\eqref{busch2} are separately minimized for the choice of $\vec{q_1}=-\vec{x}$ and ${\vec{q_2} = - \vec{z}}$, providing 
\begin{equation}
t \geq (\sqrt{2}-1)^2 \approx 0.1716.
\end{equation} 

To conclude the proof, we must prove that the use of an unbiased $N_{\pm|x}$ is optimal. Consider a biased observable $N_{+|x} = \frac{1}{2}((1+\alpha_x)\openone+\vec q_x\cdot\vec\sigma)$, then $N_{+|x} \succeq 0$ implies $|\vec{q_x}|\leq 1- |\alpha_x|$. Hence, a biased $N_{+|x}$ would decrease the length of $\vec{q_x}$ by a factor of $1- |\alpha_x|$, and consequently increase $t$ to satisfy the necessary condition of Eq.~\eqref{busch2}. In addition to that, for biased observables, Eqs.~\eqref{buschcond},\eqref{busch2} become only necessary, so even when it is satisfied, one may have to further increase $t$ to obtain joint measurability.

\bibliography{Device_indep_incomp_arXiv}


\end{document}